\documentclass[3p]{elsarticle}
\journal{Biographical Memoirs of Fellows of the Royal Society}

\usepackage{amsmath,amssymb,array,booktabs,dsfont,graphicx,mathtools,multirow,rotating,tensor,verbatim}
\usepackage[bookmarksnumbered,linktocpage,pdfstartview=FitH]{hyperref}
\usepackage[all]{hypcap}
\usepackage{graphicx}
\usepackage{caption}
\usepackage{textcomp}
\usepackage{floatrow}

\newcolumntype{M}[1]{>{$}{#1}<{$}}
\newcolumntype{B}[1]{>{\mathbf\bgroup}{#1}<{\egroup}}
\newcolumntype{K}{>{\lvert}{c}<{\rangle}}

\numberwithin{equation}{section}

\newcommand{\be}{\begin{equation}}
\newcommand{\ee}{\end{equation}}
\newcommand{\bea}{\begin{eqnarray}}
\newcommand{\eea}{\end{eqnarray}}

\begin{document}

\begin{frontmatter}
\title{
STANLEY DESER\footnote{To appear in Biographical Memoirs of Fellows of the Royal Society}
\\
~
\\
19th March 1931 - 21st April 2023
\\
~
\\
ForMemRS 2021}

%Biographical Memoirs of Fellows of The Royal Society

\author{M.~J.~Duff FRS\footnote{m.duff@imperial.ac.uk}}
\author{and K.~S.~Stelle\footnote{Kelly Stelle, my co-author, dear friend of 50 years and former student of Stanley Deser, tragically died before this memoir was completed. I dedicate it to his memory. MJD.}}

\address{The Blackett Laboratory, Imperial College London, Prince Consort Road,\\London SW7 2BW, U.K.}

\begin{abstract}
Stanley Deser was Ancell Professor Emeritus at Brandeis University and Visiting Professor at the California Institute of Technology.  He was a world-renowned theoretical physicist whose accomplishments include the Arnowitt-Deser-Misner (ADM) formulation of general relativity, supergravity with Zumino, the dynamics of flat space with Jackiw and 't Hooft and many other contributions to the quantum theory of gravity. In the 1980s he was instrumental in securing National Science Foundation approval of the LIGO gravitational wave experiment against much opposition. He has a long association with the UK, dating back to collaborations with Bondi and Pirani at King's College,  Gibbons at Cambridge and Sciama at Oxford, where he was  visiting Fellow of All Souls, as well as Duff, Isham and Stelle at King's and Imperial.  He was awarded the Einstein Medal in 2015 for important contributions to general relativity, in particular the development of the ADM formalism. Other awards include: The Dannie Heineman Prize 1994; Fellow of the National Academy of Sciences 1994; Honorary Member of the Torino Academy of Sciences 2001; Member of the American Academy of Arts and Sciences 1979, Fellow of the American Institute of Physics.\end{abstract}
\end{frontmatter}
\begin{figure*}[ht]
\includegraphics[scale=1.00]{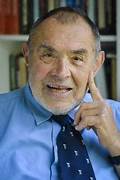}
\caption{Stanley Deser \textcopyright\  Brandeis} 
\label{Deser}                     
\end{figure*}
%\end{frontmatter}
\bigskip
\bigskip
\bigskip
\bigskip

\newpage

\section{\bf LIFE}
\label{sec:Introduction}
\subsection{\bf  Refugee}
Stanley Deser was born in Rovno, Poland, a city now known as Rivne, Ukraine, on March 19, 1931, the only child of Norman and Miriam.
 Norman was a chemist and the family were reasonably well off but fled Poland's antisemitic regime in 1935 first for Palestine and then for Paris where they lived until 1940.
 
 Stanley's formal education began in the French public and private school systems.  In his autobiography, {\it Forks in the Road} [Deser 2021], he recalls:``The kindergarten, known as
the Ecole Maternelle, was the polar opposite of its name. We were crowded, about 43--44
kids per class, all wearing black aprons, presided over by a formidable presence whose
main pleasure was using her ferrule, but not for measuring. She loved to quell any
disturbances and menaces in her class of dangerous five year-olds with it. Another wrinkle
in the system was the weekly reseating of the students according to our perceived
performance in the previous one. I always ended up around 43rd or 44th in line; I may
have reached 42nd once. This technique of course achieved the wrong end, because all we
stupid children were furthest away from the font of wisdom. Then my parents decided to put me, not that they had much choice, into a horrible little private school;
 there were two of us in my class. It wasn't really a school, just a woman and a few stray kids. At least I wasn't seated at number 44.''

 The Deser family lived on Rue Lafayette, a broad boulevard on the north side of Paris.  With World War II engulfing Europe, they began to see the exodus of Dutch and Belgian refugees.  In 1940, they too realised the danger and decided to leave everything. They owned a tiny car, a Simca 5.   Neither of his parents could drive; his uncle,
who lived with them, was the only one who knew how. Stanley wrote of his parents ``I rushed with my father to empty our safe. That evening, my mother sewed the coins into a belt of towels, a much-practiced maneouver of refugees.'' After a restless night they jumped into the car and escaped Paris hours before the Nazi tanks came rolling in.

About a week from Paris, they reached Bayonne and Biarritz, very close to the southwest
Atlantic Spanish border, having by-passed Orleans and Bordeaux.  Aristides de Sousa Mendes, the Portuguese consul in
Bordeaux who oversaw the French southwest region, foreseeing the coming slaughter,
unilaterally decided to issue transit visas to all who were waiting in his sub-consulates. The Desers put their passports into a
basket that was hauled up through a window and back down with the life-saving visas. Armed with the necessary documents, they abandoned the car and caught the last safe train out of France and secured passage on one of the last boats out of Lisbon to New York.

\subsection{\bf The New  World}

 ``In the old Ellis Island days, the immigration inspectors arbitrarily renamed the newly landed. Our experience was not so drastic; we were allowed to make our own choice of first name. I was born Salomon, then underwent a change to Lucien in France because my parents felt it would ease my way. We were told Lucien would never do in the US, so among the options we were offered, Stanley seemed the least objectionable. ''
 
Anyone with a childhood as tumultuous as his might reasonably expect adulthood to be something of an anticlimax, but for the now ``Stanley'' Deser another adventure was only just beginning, albeit an intellectual one in theoretical physics.  Readers of his autobiography  can relish the anecdotes, told with wit and literary aplomb, involving the greatest minds of 20th century physics.  They will come away with a greater appreciation of a golden era for quantum field theory and general relativity to which Deser was himself a major contributor. One such story recalls former hero of the Manhattan Project J Robert Oppenheimer, betrayed and humiliated by H-bomb advocate Edward Teller, phoning Teller to politely request he approve the young Deser's job application (which he did).

Stanley entered Brooklyn College of the City of New York with a competitive Regents Scholarship in January 1946, graduating in June 1949 BS Summa cum Laude. Professor Melba Phillips was a major influence, as was cultural life in New York City.  The next step was  Harvard University, where he gained his MA in 1950, and PhD in 1953 with Nobel laureate Julian Schwinger.  It was here he discovered his passions for theoretical high energy and particle physics.  His postdoctoral positions included the Institute for Advanced Study, Princeton 1953-55 working with M. Goldberger and W. Thirring, the Niels Bohr Institute 1955-57, where he wrote his first paper on quantum gravity, and Harvard University 1957-1958, working with W. Gilbert (FRS 1987) and E. C. G. Sudarshan.

Stanley was Married 25 August 1956 to Elsbeth Klein, an artist. He joined Brandeis University in 1958 first as  Associate then full Professor of Physics until emeritus in 2005.  From 1964 to 1970  he was Co-Director and Co-Editor, Brandeis Summer Institutes in Theoretical Physics and from 1969 to 1971 and from 1976 to 1977 he was Chairman of the Brandeis Physics Department. He held visiting professorships at  the Institut Henri Poincare of the Sorbonne;  University of the Republic, Uruguay;  Faculty of Science, University of Paris-Orsay;  Kings College, London;  College de France, Paris;  All Souls College, Oxford;  Imperial College, London; Physics Department, Texas A\&M University;  Physics Department UCLA;  Observatoire de Meudon and Institut des Hautes Etudes Scientifiques, France; University of Vienna and California Institute of Technology.   In the 1980s he was instrumental in securing NSF approval of the LIGO gravitational wave experiment against much opposition.  His numerous awards are listed in Section \ref{awards}.

 Stanley died on April 21 2023. He leaves behind three daughters, Toni, Clara, and Abigail, as well as four grandchildren, Ursula, Oscar, Louise, and Simon. He was preceded in death by his wife  Elsbeth  and a daughter, Eva.

\section{\bf WORK} 

This section reviews Stanley's central role in advancing theoretical high energy physics at a more technical level, emphasising the key turning points in the development of the discipline.
Quantum Gravity is the would-be unified theory that eluded Einstein, mystified the greatest minds of twentieth century physics and continues to baffle those of the twenty-first. Stanley Deser was not only one of the world's leading experts on this conundrum but through his work on the canonical formalism, supergravity and string theory perhaps also one of the providers of its solution.

The job of  theoretical physicists is twofold :(1) To explain the data garnered by their experimental colleagues; (2) To imagine new phenomena that no-one has yet thought of and for which, therefore, there are as yet no experiments. This is speculation but speculation in the strait-jacket imposed by mathematical consistency and agreement with what we know already to be true. Throughout his long career, Deser focussed on such foundational consistency conditions, for example with higher spins [Deser 1979,1985b], massive gravity [Deser 1972], positivity of energy [Deser 1968, 1977b] and electromagnetic duality [Deser 1976e,1977], he questioned not merely whether a theory was elegant or useful but whether it could exist at all.

\subsection{\bf The ADM formalism}
\label{ADM}

Building on earlier work of Paul Dirac (FRS 1958), the Arnowitt-Deser-Misner (ADM) formulation
 [Deser 1962] did for Einstein's general relativity what Hamilton did for classical mechanics.  The formalism regards spacetime as a family of spacelike surfaces labeled by their time coordinate $t$, and with coordinates on each slice given by $x^i$. The dynamical variables of this theory are taken to be the metric tensor of three-dimensional spatial slices $ g _{ij}(t,x^{k})$
 and their conjugate momenta.  Using these variables it is possible to define a Hamiltonian, and thereby write the equations of motion for general relativity in the form of Hamilton's equations.

Analysing  the dynamics of the gravitational field is complicated by the general coordinate invariance that underlies the theory. Ordinarily in field theory, it is suffiicient to specifiy the fields and their first time derivatives initially in order  to determine their time development.  For general relativity, however,  it is necessary that the metric field be separated into the parts carrying the true dynamical information and those parts characterizing the coordinate
system. In this respect, the coordinate invariance plays a role similar to the gauge invariance of the Maxwell field in
electromagnetism. In both cases there are redundant variables. 

ADM begin by making a $3+1$ dimensional decomposition of  the metric;
\be
g_{ij} \equiv {}^4g_{ij}, ~~~~~~~~~N \equiv(-{}^4g^{00})^{-1/2}~~~~~~~~~N_i \equiv{}^4g_{0i} \nonumber
\ee 
\be
\pi^{ij} \equiv \sqrt{-{}^4g}({}^4\Gamma{}^0{}_{pq}- g_{pq}{}^4\Gamma{}^0{}_{rs} g^{rs}) g^{ip}g^{jq}\nonumber
\ee
where every 4-dimensional quantity carries a prefix ${}^4$, thus ${}^4\Gamma{}^{\rho}_{\mu\nu}$ is the 4-dimensionsal affine connection, and all other quantities are 3-dimensional. In particular $g^{ij}$ is the inverse of $g_{ij}$. The full metric may thus be written
${}^4g_{00}=-(N^2-N_iN^i)$ where $N^i=g^{ij}N_j$,  ${}^4g^{0i}=N^i/N^2$, ${}^4g^{00}=-1/N^2$, ${}^4g^{ij}=g^{ij}-N^i N^j/N^2$ and $\sqrt{-{}^4g}=N\sqrt{g}$.

\begin{figure*}[ht]
\includegraphics[scale=1.50]{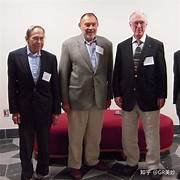}
\caption{
Richard Arnowitt, Stanley Deser and Charles Misner at the ADM-50: A Celebration of Current GR Innovation conference held in November 2009 to honor the 50th anniversary of their paper.Photographer unknown; from Wikipedia under a CC BY-SA 3.0 licence.
\textcopyright\  } 
\label{Deser}   
\end{figure*}  
The  3-dimensional variables, such as $g_{ij}$, are chosen to be three-covariant under all coordinate transformations that leave the $t=const$ surfaces unchanged. In contrast $N$ and $N_i$ are not. Momentum is more difficult to define within the surface,
since it refers to motion in time leading out of the original t = const surface. The second fundamental form does
not depend on the choice of coordinates away from the surface. This may also be seen from a
standard definition $K_{ij}=n_{({i:j})}$,
which expresses $K_{ij}$ as the covariant spatial part of the tensor $n_{(\mu;\nu)}$ (the four-dimensional covariant derivative of the unit normal,
$ n_{\mu}=-N \delta^0{}_{\mu}$ to the surface).  For convenience ADM choose  the closely related variable 
$
\pi^{ij}=\sqrt{g}(K^{ij}-g^{ij} K)\nonumber
$. 
 While the 12 $(g_{ij}, \pi^{ij})$ constitute a complete set of Cauchy data, they do not provide a minimal set.
We may eliminate four by using the constraint equations and then another 4 by imposing coordinate  conditions. This leaves two degrees of freedom which was to expected since a massless spin 2 field has just two.

Introducing the transverse traceless components $\pi^{ij}{}^{TT}$ and $g_{ij}{}^{TT}$ where, for a symmetric second rank tensor $f_{ij}$
\[
f_{ij}{}^{TT},{}_j\equiv 0~~~~~f_{ii}{}^{TT}\equiv 0\nonumber~~~~~~~ f^T\equiv f _{,ii}-\frac{1}{\nabla^2}f_{ij,ij}
\]
ADM were able to write down the fundamental equal time Poisson bracket relations for $\pi^{ij}{}^{TT}$ and $g_{ij}{}^{TT}$
\[
[g_{ij}{}^{TT}({\bf x}),\pi^{kl}{}^{TT}({\bf x'})]=\delta_{ij}{}^{kl}({\bf x}-{\bf x'})
\]
\[
[g_{ij}{}^{TT}({\bf x}),g_{kl}{}^{TT}({\bf x'})]=0
\]
\[
[\pi^{ij}{}^{TT}({\bf x}),\pi^{kl}{}^{TT}({\bf x'})]=0 
\]
where $\delta_{ij}{}^{kl}({\bf x}-{\bf x'}) $ is a conventional Dirac $\delta$ function, symmetric, transverse and traceless on each pair of indices.

In summary,  by casting general relativity into canonical form, the Arnowitt-Deser-Misner approach formed the basis of early attempts to quantize gravity and continues to be used in modern methods based on supergravity, extra dimensions, string theory and M-theory. 

\subsection{\bf Non-renormalizablity}
\label{renorm}

 One of the major obstacles facing quantum gravity were the infinities that rendered the theory ``non-renormalizable'', in contrast to those arising in models of the strong nuclear, electromagnetic and weak nuclear forces, where they could be absorbed into a renormalization of the masses and coupling constants. The arrival of dimensional regularization in 1972 opened the door to loop corrections to Feynman diagrams  in quantum field theory which manifestly preserved gauge invariance. In particular, it was also applied by 't Hooft and Veltman [`t Hooft 1972] to quantum gravity 
showing that although pure gravity was finite at one loop, it ceased to be finite when coupled to a 
scalar  field. This raised the question of whether couplings to other forms of matter might avoid this problem.

For example, let us consider the case of the Einstein-Maxwell Lagrangian 
\be
L=-\sqrt{-g}[\kappa^{-2}R+\frac{1}{4}F_{\mu\nu}F^{\rho\sigma}g^{\mu\rho}g^{\nu\sigma}]\nonumber
\ee
where $\kappa^2=8\pi G$ and $G$ is Newton's constant, $g_{\mu\nu}$ is the metric tensor,  $g^{\mu\nu}$ is its inverse and $g$ its determinant. $R$ is the curvature scalar and $F_{\mu\nu}$ is the curl of the Maxwell potential $A_{\mu}$, 
$
F_{\mu\nu}=\partial_{\mu}A_{\nu}-\partial_{\nu}A_{\mu}.
$
Deser and Van Nieuwenhuizen [Deser 1976a] calculated the one-loop counterterms $\Delta L$ necessary to cancel the infinities using dimensional regularization with parameter $\epsilon=D-4$ where $D $ is the spacetime dimension. Since $\kappa$ is the only dimensional constant there will be no divergences proportional to $L $ itself nor a cosmological term $\sqrt{-g}$.

Explicit calculation  yields one non-vanishing term
\be
\Delta L=(\sqrt{-g}/ \epsilon)\frac{137}{60}R_{\mu\nu}{}R^{\mu\nu}\nonumber
\ee
and the theory is not one-loop renormalizable. Similar negative results  were found for Yang-Mills fields [Deser 1974b, 1974c] and for spin 1/2 fermions [Deser 1974d], but as discussed in Section \ref{infinite} the game
was about to change dramatically with the arrival of Supergravity.
 \subsection{\bf Weyl anomalies}
\label{Weyl} 
The conformal invariance under Weyl rescalings of the metric tensor 
\be
g_{\mu\nu}(x) \rightarrow \Omega^2(x)g_{\mu\nu}(x)\nonumber
\ee
displayed by classical massless field systems in interaction with gravity no longer survives in the quantum theory.   As a consequence, the regularized energy-momentum tensor develops a non-vanishing trace:
\be
g^{\alpha\beta}<T_{\alpha\beta} > \neq 0 \nonumber
\ee
In April 1976 Deser, Duff and Isham wrote down the most general form of the trace of the energy-momentum tensor in various
dimensions [Deser 1974a]. By general covariance and dimensional analysis, it must take
the following form:

\bea
D =2 \nonumber ~~~~~~~~~~~~~~~~~~~~~~~~~~~~~~~~~~~~~~~~~
g^{\alpha\beta}<T_{\alpha\beta} >&=& cR\\\ \nonumber
D =4~~~~~~~~~~~~~~~~~~~~~~~~~~~~~~~~~~~~~~~~~
g^{\alpha\beta}<T_{\alpha\beta}>&= &\alpha R^2 + \beta R_{\mu\nu}R^{\mu\nu}+
\gamma R_{\mu\nu\rho\sigma}R^{\mu\nu\rho\sigma} +\delta {\Box R}\\\ \\\ \nonumber
D =2n~~~~~~~~~~~~~~~~~~~~~~~~~~~~~~~~~~~~~~~
g^{\alpha\beta}<T_{\alpha\beta} >&= &(Riemann)^n+...\\\  \nonumber
\eea
(At one-loop, and ignoring boundary terms,
there is no anomaly for D odd).
The significance of this paper was to demonstrate
that, in addition to scheme-dependent terms such as $\Box R$ which can be removed by the addition of finite local counterterms such as $R^2$, there are scheme-independent terms such as $ \alpha R^2 + \beta R_{\mu\nu}R^{\mu\nu}+\gamma R_{\mu\nu\rho\sigma}R^{\mu\nu\rho\sigma}$ which cannot, thus laying to rest any lingering doubts about the inevitability of Weyl
anomalies. As is well-known, the effective action arising from closed loops
of massless particles is non-local. The title of the paper {\it Non-local Conformal
Anomalies} was chosen to emphasize that although the trace of the stress tensor and infinite counterterms are local, for example 
$C_{\mu\nu\rho\sigma}C^{\mu\nu\rho\sigma}$ where 
\[
C_{\mu\nu\rho\sigma}=R_{\mu\nu\rho\sigma}+\frac{1}{2}(R_{\mu\rho}g_{\nu\sigma}-R_{\nu\rho}g_{\mu\sigma}+R_{\nu\sigma}g_{\mu\rho}-R_{\mu\sigma}g_{\nu\rho})+\frac{1}{6}R(g_{\nu\rho}g_{\mu\sigma}-g_{\mu\rho}g_{\nu\sigma})
\]
is the Weyl tensor, the finite effective action responsible for the $ (curvature)^2$ anomalies is not, for example
\[
C_{\mu\nu\rho\sigma}~ ln ~\Box~ C^{\mu\nu\rho\sigma}
\]
This means, incidentally, that Starobinsky's original model of cosmic inflation [Starobinsky 1980]
which was driven by the Weyl anomaly, involved a non-local Lagrangian and not the local
$R + R^2$ as is often stated.  One may construct a local action involving extra scalar fields which yields the non-local action after integrating out the
scalars [Riegert 1984]. However, the exact form of the non-localities is still a matter of
debate. 
In  September 1976 Eguchi and Freund [Eguchi 1976] considered two topological
invariants: the Pontryagin number 
\[
P=\frac{1}{(4\pi)^2} \int d^4x {\sqrt g} \epsilon^{\mu\nu\rho\sigma} R_{\mu\nu}{}^{\alpha\beta}R_{\rho\sigma}{}_{\alpha\beta}
\]
and the Euler number
\[
\chi=\frac{1}{2(4\pi)^2} \int d^4x {\sqrt g}[R_{\alpha\beta\gamma\delta}{}R^{\alpha\beta\gamma\delta} -4R_{\mu\nu}{}R^{\mu\nu}+R^2]
\]
and posed the question: to what anomalies do they contribute? They found that P appears in the integrated divergence of the axial
fermion number current, thus providing a novel topological interpretation of
the anomaly found by Kimura [Kimura 1969] in 1969 and Delbourgo and Salam [Delbourgo 1972] in 1972.
However, they found no analogous role for $\chi$. Drawing on [Deser 1976a], however, one could show that for Conformal Field Theories (CFTs) in $D=4$, the trace of
the stress tensor depends on just two constants  [Duff 1977,1994]:
\[
g^{\alpha\beta}<T_{\alpha\beta} >=\frac{1}{(4\pi)^2}(cF-aG)
\label{canda}
\]
where F is the square of the Weyl tensor 
$F=C_{\mu\nu\rho\sigma}C^{\mu\nu\rho\sigma}
$
and G is the Euler number density
$
(4\pi)^2\chi=\int d^4x{\sqrt g}G. 
$
For free CFTs with $N_s$ massless fields of spin $s$
\[
720c = 6N_0 + 18N_1/2 + 72N_1~~~~~~~~~~~~ 720a = 2N_0 + 11N_1/2 + 124N_1
\]
Note the inequalities on the ratio a/c: $31/16 \geq a/c\geq 1/3$ where the upper and lower bounds are saturated by a single vector and a
single scalar, respectively.  Remarkably, these bounds continue to hold true when the CFT is interacting.

 The ``$ conformal $-$ Euler $  '' structure of the anomaly, $cF-aG$, was generalized to arbitrary even dimensions by Deser and Schwimmer [Deser 1993]
 \[
 g^{\alpha\beta}<T_{\alpha\beta}> =\sum_i c^iI_i-(-1)^{D/2}aE_D
\label{canda2}
\]
where $E_D$ is the Euler density arising from scale-free contributions to
the effective gravitational action and the $I_i$ are local conformal scalar
polynomials involving powers of the Weyl tensor and its derivatives;
their number increases rapidly with dimension. The two terms are called Type B and Type A, respectively.

The two dimensional anomaly is purely Type A, suggesting that the  c-theorem on renormalisation group flow in D = 2
[Zamalodchikov 1986] would generalize to an a-theorem in D = 4.  This has
now been proved [Komargodski 2011].
These theorems confirm that theories at high energies have more degrees of freedom than theories at low energies and that information is lost as we flow from the former to the latter.

Spacetime Weyl anomalies have found application in quantum corrections to the Schwarzschild solution and Newton's law, particle creation, the Hawking effect, inflationary cosmology,
asymptotic safety, wormholes, viscosity
bounds, condensed matter physics, hydrodynamics, the
graviton mass in the braneworld, conformal collider physics,
quantum entanglement, log corrections to black hole
entropy, generalized mirror symmetry, and doublecopy theories. An especially important role is played by the holographic Weyl anomaly in the AdS/CFT correspondence [Henningson 1998, Graham 1999]

The cancellation of worldsheet Weyl anomalies not only determines
the critical dimensions D = 26 for strings and D = 10 for superstrings
 but also provides the derivation of the spacetime Einstein
equations [Callan 1980].

\subsection{\bf Consistent Supergravity}
\label{super}
The discovery of Supergravity, a supersymmetric version of general relativity, by Freedman, Ferrara and Van Nieuwenhuizen  [Freedman 1976] and  by Deser and Zumino [Deser 1976c] represents an important milestone in theoretical physics. If Einstein had not already discovered general relativity, supersymmetry would have forced us to invent it:  a local supersymmetry transformation is the square root of a general co-ordinate transformation, the basis of relativity. 
The emphasis of the Deser-Zumino paper was to find a combined spin 2 - spin 3/2
extension of general relativity is which is free of the
usual higher spin inconsistencies. This was most easily achieved in first-order formalism where the vierbein $e_{\mu}{}^a$ and spin connection $\omega_{\mu}{}^{ab}$ are treated independently. The result was a Lagrangian which was the sum of an Einstein term and a minimally coupled Majorana Rarita-Schwinger field $\psi_{\mu}$.

\be
L=-\frac{1}{2}eR-\frac{1}{2}\epsilon^{\mu\nu\rho\sigma}{\bar \psi_{\mu}}\gamma_5 \gamma_{\nu}D_{\rho}\psi_{\sigma}\nonumber
\ee
where $e=det ~e_{\mu}{}^a$, $R=e^{\mu}{}_a e^{\nu}{}_b R_{\mu\nu}{}^{ab}$ and where
$D_{\mu}=\partial_{\mu}-\omega_{\mu}{}[\gamma_a,\gamma_b]^{ab}/8$
satisfies $[D_{\mu},D_{\nu}]=-R_{\mu\nu}{}^{ab}[\gamma_a,\gamma_b]/8$
Varying the gravitino $\psi_{\mu}$, the spin connection $\omega_{\mu}{}^{ab}$ and the vierbein $e_{\mu}{}^a$ independently, the equations of motion are
\bea
R^{\lambda} \equiv \epsilon^{\lambda\mu\nu\rho} (\gamma_{\mu} D_{\nu }\psi_{\rho}
-\frac{1}{4}\gamma_{\tau} C_{\mu\nu}{}^{\tau}\psi_{\rho})&=&0 \nonumber\\
 C_{\mu\nu}{}^{\tau}&=&\frac{i}{2}{\bar \psi}_{\mu}\gamma^{\tau}\psi_{\nu} \nonumber\\
G^{\tau\mu}&=&\frac{i}{2}\epsilon^{\lambda\mu\nu\rho} {\bar \psi}_{\lambda}\gamma_5 \gamma^{\tau}D_{\nu}\psi_{\rho} \nonumber
\eea
where $G^{\tau}{}_a =R^{\tau}{}_a -e^{\tau}{}_aR/2$ with $R_{\tau a}=R_{\tau \lambda a}{}^{\lambda}$ is the (non-symmetric) Einstein tensor and
$C_{\mu\nu}{}^{a}=D_{\mu}e_{\nu}{}^a-D_{\nu}e_{\mu}{}^a$
is the torsion.  In general, the consistency problem arises when applying $D_{\lambda}$ to the left hand side of the Rarita-Schwinger equation one obtains an undesirable constraint on the solutions. In the Deser-Zumino set-up, however, $D_{\lambda}R^{\lambda}$ vanishes as a consequence of the other equations of motion.

Supergravity soon became a hot topic for research. For a review, see [Van Nieuwenhuizen 1981].
It was found that the $N=1$ supermultiplet with one spin 2 and one spin 3/2 can be enlarged to $N=2$ with one spin 2, two spin-3/2 and one spin 1 and so on to a maximal $N=8$  with one spin 2, eight spin 3/2, twenty-eight  spin 1; fifty-six spin 1/2 and seventy spin 0.  Going beyond $N=8$ leads to inconsistent higher spins. Since each supersymmetry is associated with a 4-component spinor parameter, this means a maximum of $ Q=32 $ components. If one asks what is the maximum dimension of spacetime, $D$, admitting $ Q=32$, one finds $D=11$. Thus supergravity led a revival of the Kaluza-Klein idea of extra dimensions which was incorporated in superstring theory whose critical dimension is $D=10$ and M-theory whose critical dimension is $D=11$.

Deser and Zumino continued their collaboration with influential papers on the spinning string theory [Deser 1976d] and broken supersymmetry [Deser 1977b].

\subsection{\bf Infinities in supergravity}
\label{infinite}
One of the chief questions in the construction of supergravity [Freedman 1976, Deser 1976c] was whether
this theory would prove renormalizable, in contrast to ordinary Einstein gravity with or without coupling to  spin 0, 1/2 and 1,  as discussed in Section 2.2. Supersymmetry places extra constraints on the possible counterterms.  Indeed, pure $N=1$ supergravity (gravity plus spin 3/2 gravitino) was shown to be finite at one and two loops as was $N=2$ (gravity plus 2 gravitinos plus a photon) [Grisaru 1976].

However, hopes that these symmetry arguments would continue to three loops and beyond were dashed when Deser, Kay and Stelle [Deser 1977a ]  showed that a candidate counterterm exists whose bosonic part is given by the square of Bel-Robinson tensor $T_{\mu\nu\alpha\beta}$
\be
T_{\mu\nu\alpha\beta}=-R^{\lambda}{}_{\alpha}{}^{\rho}{}_{\mu}R_{\lambda\beta\rho\nu}+*R^{\lambda}{}_{\alpha}{}^{\rho}{}_{\mu}*R_{\lambda\beta\rho\nu}\nonumber
 \ee
 which is totally symmetric, traceless and conserved on shell.  The study of ultraviolet divergences in supergravity, of which Stanley Deser was a pioneer, has since become a major industry. But just because a candidate counterterm exists it does not necessarily mean it will actually appear and supergravity turns out to exhibit surprising degrees of finiteness, which are still unexplained. Here we reproduce a summary of Bern et al [Bern 2023] in Table 1.
 
 It is still an open question whether the maximally supersymmetric $(N=8, D=4)$ supergravity is or is not finite.

\begin{table}
$\begin{tabular}{|r|r|r|r|r|r|} \hline
D&Q&N&Counterterm&Loop&Divergence?\\ \hline
4&32&8&$D^8R^4$&7&unknown\\ \hline
4&16&4&$R^4$&3&no\\ \hline
4&20&5&$D^2R^4$&4&no\\ \hline
5&16&4&$R^4$&2&no\\ \hline
\end{tabular}$
\caption{Counterterms corresponding to the first potential divergence that satisfy all proven symmetry constraints, but which nevertheless fail to appear.  D denotes the spacetime dimension, Q the number of supercharges and N the number of supersymmetries.}
\end{table}
\subsection{\bf  Topologically Massive Gravity}

Another important strand of Deser's work concerned the dynamics of gravity and gauge theories in lower-dimensional spacetimes. In an  influential collaborations with Jackiw and Templeton (garnering over 3000 citations), vector and gravity models were studied in three-dimensional space-time, where novel, gauge invariant, terms of topological origin give rise to masses for the gauge fields [Deser 1982].

In separate work with  Jackiw and 't Hooft, Deser investigated the peculiar dynamics of three dimensional gravity [Deser 1984]. By virtue of the identity,
\[
R^{\mu\nu}{}_{\alpha\beta}=\epsilon^{\mu\nu\sigma}\epsilon_{\alpha\beta\gamma}R^{\lambda}{}_{\sigma}
\]
%G^{\lambda}{}_ {\sigma}=0$ 
linking curvature and Einstein tensors, empty regions where $G^{\lambda}{}_ {\sigma}=0$ are flat while internal ones with non-vanishing stress-tensors $G^{\lambda}{}_ {\sigma}=8\pi GT^{\lambda}{}_ {\sigma}$ are not.
\subsection{\bf Legacy}

Given Stanley Deser's pioneering contributions, it is worthwhile reviewing the current status of 
 supergravity [Freedman 1976,Deser 1976c] and supersymmetry in general.

Supersymmetry unites bosons (particles with integer spin ) and fermions (particles with odd half-integer spin). Roughly speaking, if our equations remain the same when we swap the bosons and fermions, we say they are supersymmetric.  For example, known bosons: spin-0 Higgs, spin-1 photons and spin-2 gravitons have as their superpartners the not-yet-discovered fermions: spin-1/2 Higgsinos, spin-1/2 photinos and spin-3/2 gravitinos. Whereas known fermions: spin-1/2 electrons and spin-1/2 quarks have as their superpartners the not-yet-discovered bosons: spin-0 selectrons and spin-0 squarks. The failure to detect these supersymmetric particles at the Large Hadron Collider (LHC) in Geneva has affected the theoretical  physics communities in different ways.

 In the 1970s the formal-theory/quantum-gravity community  embraced global and local supersymmetry as fermionic extensions of Einstein's Special and General Relativities, respectively, providing a natural framework for unifying gravity with the other forces.  This continued with the arrival of the 1984 string revolution which replaced  0-dimensional particles by 1-dimensional strings as the fundamental building blocks, thereby solving the problem of ultraviolet divergences. Supergravity was seen to be the low-energy limit of superstring theory. In the 1995 membrane revolution, string theory was seen to be the weak-coupling limit of a deeper, more profound, theory called M-theory which unified the five consistent  $D=10$ string theories and $D=11$ supergravity by incorporating extended membrane-like objects called $p$-branes of which strings and particles were respectively the $p=1$ and $p=0$ special cases. $M$-theory owes as much to supergravity as it does to string theory especially in the context of $11$  dimensions where there are no superstrings.

In the meantime, however,  the particle physics community became enamoured of supersymmetry for its own purposes, namely as a possible solution to the ``gauge hierarchy problem'' (the huge disparity between the energy scale of the weak nuclear force and that of gravity, the Planck scale). Omitting the details, if supersymmetry were to be the solution to the gauge hierarchy problem, one would expect to see the missing superpartners at the Large Hadron Collider; if not then one would not because there would be no reason for them to be light enough. So the failure to detect at them does not rule out supersymmetry itself and the quantum gravity community, many of whom never endorsed the hierarchy idea in the first place, were mainly unfazed.

We recall that there were three outstanding issues in quantum gravity in the 1970s: (1) Ultraviolet divergences and non-renormalizability; (2) The microscopic origin of the Bekenstein-Hawking black hole entropy; (3) The black hole information paradox (the apparent clash between quantum mechanics and general relativity when a black hole evaporates).
Fifty years on, a glance at  {\it hep-th}  https://arxiv.org/list/hep-th/new confirms that supergravity, string theory and M-theory continue to be the favourite candidates for reconciling gravity and quantum mechanics. This is not without good reason.  Supersymmetry in the form of string theory provided an answer to problem (1), supersymmetry in the form of M-theory has provided an answer to problem (2) and supersymmetry in the form of AdS/CFT has (according to Hawking and others) provided an answer to (3). Moreover, no other theory provides adequate answers to any of (1), (2) or (3). Supersymmetry is alive and kicking and Supergravity continues to be an active area of research both as the low energy limit of string theory and in its own right, thanks in large part to Stanley Deser.

More on Stanley Deser's scientific contributions may be found in the Proceedings of the Deserfest [Liu 2006], a tribute to S. Deser [Tekin 2023], the Stanley Deser Memorial Volume [Henneaux 2025] in particular [Duff 2025].

 \section{\bf PERSONAL RECOLLECTIONS OF MJD}

I first came into contact with Stanley as a student at the last (1970) Brandeis Summer School, of which he was a co-organizer. He was easy to spot because there was always a cloud of pipe-smoke following him around, but did not get to know him personally until he visited King's College London in 1976 where he was a guest of Chris Isham and I was a postdoc. 

In fact he had a long association with the UK, dating back to collaborations with Hermann Bondi (FRS 1959) and Felix Pirani at King's, with Dennis Sciama (FRS 1983) at Oxford where he was visiting Fellow of All Souls and Gary Gibbons at Cambridge. The combination of permanent staff, postdocs and visitors made King's a centre of activity for classical gravity (Felix Pirani and David Robinson), quantum field theory in curved spacetime (Tim Bunch, Steve Christensen, Paul  Davies, Larry Ford, Steve Fulling and Bill Unruh (FRS 2001) and quantum gravity (Stanley Deser, Chris Isham, John G. Taylor and the authors MJD and KSS).  Many appear in the photograph Fig.3, courtesy of Steve Christensen.  See [Robinson 2018] for a history.  But the visitor who had the greatest impact on Chris and me was Stanley (the one smoking the cigar): non-local conformal anomalies [Deser 1976a], discussed in Section \ref{Weyl}, solitons of chiral theories in three space dimensions [Deser 1976b]. The DDI collaboration continued at Imperial College with a paper on gravitationally induced CP effects via a $\epsilon^{\mu\nu\rho\sigma}R_{\mu\nu ab}R_{\rho\sigma}{}^{ab}$ in the action [Deser 1980]. See also my tribute to Chris Isham [Duff 2021].

 The chiral solitons paper introduced fourth order terms in the Lagrangian necessary to circumvent the Derrick no-go theorem in three dimensions. An alternative circumvention was to raise the sigma-model Lagrangian to the power 3/2.  We obtained what we considered to be nice results [Deser 1976b,1976c] only to discover that some had been anticipated by Skyrme in the early 60s, although this is overlooked in the received history of the Skrymion revival [Aitchison 2019].

 \begin{figure}\label{fig:p4}  
\begin{center}
\includegraphics[scale=0.7]{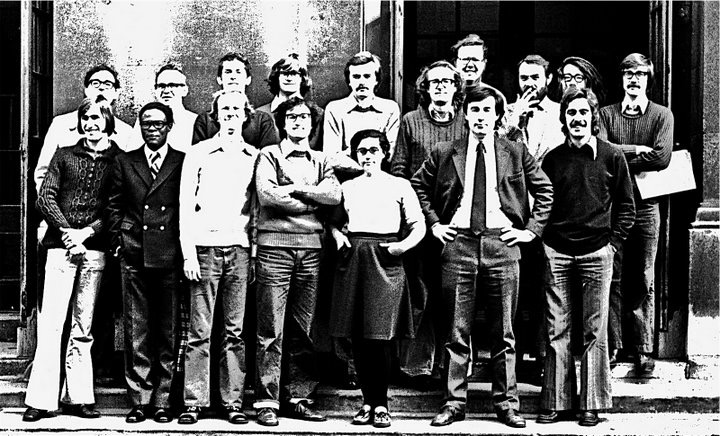}
\caption{\footnotesize{King's College London 1976. FRONT ROW: Dave Storey, Unknown, Steve Christensen, Bernard Kay, Joyce El Eini, Chris Isham, Mike Duff. BACK ROW: John Taylor, Steve Fulling, Stephen Downs-Martin, Tim Bunch, Chris Mannion, David Robinson, Bob Davis, Stanley Deser, Tezt Yoshimura, Paul Davies.. Photograph provided by Steve Christensen. }}
\end{center}
\end{figure}

My interactions with Stanley were set to increase when I accepted an offer of a two-year NSF postdoc at Brandeis, where there was very active group including Marc Grisaru, Howie Schnitzer and Burt Ovrut. I also  enjoyed the warm hospitality extended by Stanley and his wife Elsbeth at their home in Newton. Elsbeth was an accomplished artist and daughter of the Swedish physicist Oskar Klein, whose work on extra dimensions was to have a profound effect on the my career.

 \begin{figure*}[ht]
\includegraphics[scale=1.50]{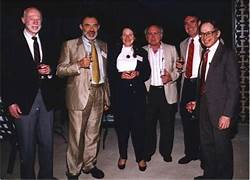}
\caption{Stanley and other quantum gravity enthusiasts celebrating the Arnowitt?Deser?Misner (ADM) formulation at Texas A\&M University. (From left to right) Bryce DeWitt, Stanley Deser, Mary K. Gaillard, Bruno Zumino, Mic\textcopyright\  Photograph by Roland Allen. } 
\label{Deser}                     
\end{figure*}

 Memories of Stanley include a stay in the Academy of Sciences Hotel for
one of the Moscow Quantum Gravity conferences circa 1980. Stanley and I set out one
night for dinner at a nearby restaurant. In those bleak years before Glasnost it was
not uncommon to be turned away from half-empty restaurants on the grounds that they
were ``full''. This is exactly what happened, but Stanley Deser would not take no
for an answer and stuck his foot in the door as the proprietor tried to close it in our
faces. Fisticuffs broke out until I managed to pull Stanley away arguing that the meal
was probably not worth getting beaten up for.

Stanley passed away in 2023 but in the later stages of his illness we kept up a vigorous
exchange of e-mails. Although those who have corresponded with Stanley will recall that
his messages, often crammed into the subject line, sometimes needed an Enigma Machine
to decipher his cryptic comments. Scientifically, Stanley's intellect was formidable. But he was 
also a cultured polymath and linguist, with a deep appreciation of literature and the arts and a
connoisseur of fine food and wine. ``Frivolous '' is the last adjective that springs to mind
when recalling Stanley Deser. All the more enjoyable therefore was the game we played:
Physics-based Horror Movies. This was prompted by an invitation I received from Peter
Freund to deliver the 9th annual Schrodinger lecture in Transylvania.  Some weak attempts included {\it Frank Einstein}, {\it Bram Stoker's Theorem}, {\it The Texas Chain Rule Massacre}, {\it Mary Shelley Glashow}, {\it Fadeev-Popov Ghosts}, {\it Paranormal Coordinates}, {\it The Killing Room}, {\it Maxwell's Demon}, {\it Phantom Black Holes} and {\it The Bourne Approximation}, but Stanley trumped them all with {\it Diracula}!

\section{\bf HONOURS AND AWARDS}
\label{awards}

%New York State Regents' Scholarship 1946-49 \medskip

Sigma Xi 1949\medskip

%AEC and NSF predoctoral Fellowships 1951-53 \medskip

%Jewett Postdoctoral Fellow 1953-55\medskip

%NSF Postdoctoral Fellow 1955-57\medskip

%Board of Editors, Journal of Mathematical Physics 1961-64, 1974-76, 1989-91, 2002- \medskip

Guggenheim Fellowship 1966-67\medskip

%Executive Committee, Division of Particles and Fields, Amer. Phys. Soc. 1970-72 \medskip

Fulbright Fellowship, Uruguay 1970\medskip

Fulbright Fellowship, France 1971-72\medskip

Distinguished Alumnus Award of Honor, Brooklyn College 1971\medskip

Fellow, American Physical Society 1972\medskip

Bude Medal, College de France 1976\medskip

Honorary D. Phil., University of Stockholm Centennial 1978\medskip

Fellow, American Academy of Arts and Sciences 1979\medskip

Ryan Memorial lecturer, Dublin Institute for Advanced Studies 1982\medskip

%Physics Advisory Committee, National Science Foundation 1982-85, 1986\medskip

Investigador Titular Ad Honorem, CIDA, Venezuela 1983\medskip

%Dannie Heineman Prize Committee, American Physical Society 1987-88, 1997-98 (Chair),
%2006-08\medskip

%Board of Editors, Classical and Quantum Gravity 1991-94\medskip

%Board of Editors, The Physical Review 1991-1993\medskip

%NSF Selection Committee for Presidential Young Investigators Awards 1991\medskip

%Board of Editors, Annals of Physics 1991-\medskip

%Scientific Advisory Committee, Institut des Hautes Etudes Scientifiques 1991-1997 \medskip

%Board of Editors, Journal of Geometry and Physics 1991-\medskip

%Scientific Advisory Committee, Institute for Theoretical Physics 1991?-995\medskip

%Steering Committee 1992-1995; Chairman 1993-94\medskip

Dannie Heineman Prize in Mathematical Physics 1994\medskip

Fellow, National Academy of Sciences 1994\medskip

%International Advisory Editorial Board, Journal of High Energy Physics 1997-\medskip

 %International Advisory Committee, Strings 2000\medskip
 
Honorary Foreign Member, Academy of Science of Torino 2001 \medskip

Honorary D. Tech., Chalmers University of Technology 2001\medskip

 Lifetime Achievement Award, Brooklyn College 2004\medskip
 
W. Bonnor Prize, GRG Society 2007\medskip

%Referee Award, American Physical Society 2008 \medskip

%ADM-fest Honoree, 2010\medskip

Einstein Medal, General Relativity Centennial, 2015\medskip

Foreign Member of the Royal Society 2021  \medskip
%\newpage

\section{\bf REFERENCES TO OTHER AUTHORS}

\noindent
[Dirac  1958]  P.A.M. Dirac, The theory of gravitation in Hamiltonian form. Proc R Soc London, Ser A. 1958;246:333-343. 
doi: 10.1098/rspa.1958.0142. \medskip

\noindent
[Kimura 1969]
T. Kimura, Divergence of Axial-Vector Current in the Gravitational
Field, Prog.Theo.Phys 42 (1969) 1192-1205.
\medskip

\noindent
[Delbourgo 1972]
 R. Delbourgo and A. Salam, The gravitational correction to PCAC,
Phys. Lett. B40 (1972) 381-382.\medskip

\noindent
[`t Hooft 1972]
G. `t Hooft and M. Veltman, Regularization and Renormalization of
Gauge Fields, Nucl. Phys. B44 (1972) 18.\medskip

\noindent
[Eguchi 1076]
 T. Eguchi and P. G. O. Freund, 
 Quantum Gravity and World Topology, 
 Phys. Rev. Lett. 37 (1976) 1251.
\medskip

\noindent
[Freedman 1976]
D. Z. Freedman, P. van Nieuwenhuizen, and S. Ferrara, Progress Toward a
Theory of Supergravity, Phys. Rev. D 13 (1976) 3214?3218.\medskip

\noindent
[Van Nieuwenhuizen 1981]
P. Van Nieuwenhuizen
Supergravity, Phys.Rept. 68 (1981)\medskip

\noindent
[Grisaru 1976]
M. T. Grisaru, P. van Nieuwenhuizen  and J. A. M. Vermasreren, 
One-Loop Renormalizability of Pure Supergravity and of Maxwell-Einstein Theory
in Extended Supergravity, PRL 37 35 1662 (1976). \medskip

\noindent
[Duff 1977]
M. J. Duff, Observations on Conformal Anomalies, Nucl. Phys.
B125 (1977) 334\medskip

\noindent
[Starobinsky 1980]
A. A. Starobinsky, A new type of isotropic cosmological models
without singularity, Phys. Lett. B91 (1980) 99?102.\medskip

\noindent
[Riegert 1984] 
R. J. Riegert,  A Nonlocal Action for the Trace Anomaly, Phys. Lett.
134B (1984) 56?60.\medskip

\noindent
[Callan 1985]
J. Callan, Curtis G., E. J. Martinec, M. J. Perry, and D. Friedan,
Strings in Background Fields, Nucl. Phys. B262 (1985) 593.\medskip

\noindent
[Zamalodchikov 1986]
 A. B. Zamolodchikov, Irreversibility of the Flux of the
Renormalization Group in a 2D Field Theory, JETP Lett. 43 (1986)
730?732.\medskip

\noindent
[Duff 1994]
M.J. Duff,
Twenty years of the Weyl anomaly,
Class.Quant.Grav. 11 (1994) 1387-1404 ,  e-Print: hep-th/9308075 [hep-th]
\medskip

\noindent
[Henningson 1998]
 M. Henningson and K. Skenderis, The holographic Weyl anomaly,
JHEP 07 (1998) 023, arXiv:hep-th/9806087.\medskip

\noindent
[Graham 1999]
C. R. Graham and E. Witten, Conformal anomaly of submanifold
observables in AdS/CFT correspondence, Nucl. Phys. B546 (1999)
52?64, arXiv:hep-th/9901021.
\medskip

\noindent
[Aitchison 2001]
 I. J. R. Aitchison, Tony Skyrme and the Origins of Skyrmions,
arXiv:2001.09944 [physics.hist-ph].)\medskip

\noindent
[Liu 2006]
J.T. Liu, M.J. Duff,  K.S. Stelle, R.P. Woodard (eds) Deserfest: A Celebration of the Life and Works of Stanley Deser.
Proceedings, Meeting, Ann Arbor, USA, April 3-5, 2004 World Scientific {2006).\medskip

\noindent 
[Komargodski 2011]
Z. Komargodski and A. Schwimmer, On Renormalization Group
Flows in Four Dimensions, JHEP 12 (2011) 099, arXiv:1107.3987\medskip

\noindent
[Robinson 2019] 
D. C. Robinson, Gravitation and general relativity at King's College London,
Eur. Phys. J. H 44 181-270 (2019)
.\medskip

\noindent
[Tekin 2019]
Bayram Tekin,
A Tribute to S. Deser: Conserved Quantities in Generic Gravity Theories
https://doi.org/10.48550/arXiv.2307.12758
\medskip

\noindent
[Duff 2021]
M.J. Duff,
Chris Isham: mentor, colleague, friend
e-Print: 2112.13722 [physics.hist-ph]\medskip

\noindent
[Bern 2023]
Zvi Bern, John Joseph M. Carrasco, Marco Chiodaroli, Henrik Johansson, Radu Roiban,
Supergravity amplitudes, the double copy and ultraviolet behavior,
arXiv:2304.07392v1 [hep-th] 
https://doi.org/10.48550/arXiv.2304.07392 46 pages, 
invited chapter for Handbook of Quantum Gravity
 (Eds. C. Bambi, L. Modesto and I.L. Shapiro, Springer Singapore.
\medskip

\noindent
[Henneaux 2025]
Marc Henneaux, Rafael I. Nepomechie Domenico Seminara,
Preface to fields, gravity, strings and beyond: in memory of Stanley Deser
 J. Phys. A: Math. Theor. 58 410301 (2025)\medskip
 
\noindent
[Duff 2025]
M.J. Duff,
Montonen-Olive duality of gauged supergravity?
J.Phys.A 58 (2025) 5, 055401
e-Print: 2410.03875 [hep-th]
DOI: 10.1088/1751-8121/ada745
Report number: Imperial/TP/2024/mjd/1\medskip

\noindent
\section{BIBLIOGRAPHY}
\noindent
\noindent
[Deser 1962]
The Dynamics of General Relativity (with R. Arnowitt and C.W. Misner) in Gravitation: An Introduction to Current Research, (ed. L. Witten), Wiley, NY (1962); reprinted as gr-qc/0405109. Reprinted in Gen.Rel.Grav. 40 (2008) 1997. \medskip

\noindent
[Deser 1968]
Positive-Definiteness of Gravitational Field Energy (with D. Brill), Phys. Rev Lett. 20, 75
(1968).\medskip

\noindent
[Deser  1972] 
Can gravitation have a finite range? 
(with D.G. Boulware)
Phys.Rev. D6 (1972) 3368 \medskip

\noindent
[Deser  1974a]
 One Loop Divergences of Quantized Einstein-Maxwell Fields, ( with P. van Nieuwenhuizen), Phys.Rev. D10 (1974) 401  \medskip

\noindent
[Deser 1974b]
Nonrenormalizability of Einstein-Yang-Mills Interactions at the One-Loop Level (with H.-S. Tsao and P. van Nieuwenhuizen), Phys. Lett. 50B, 491 (1974).\medskip

\noindent
[Deser 1974c]
One-Loop Divergences of the Einstein-Yang-Mills System (with P. van Nieuwenhuizen and H.-S. Tsao), Phys. Rev. D10, 3337 (1974).\medskip 

\noindent
[Deser 1974d]
Nonrenormalizability of Quantized Fermion Gravitation Interactions (with P. van Nieuwenhuizen), Nuovo Cimento Lett. 2, 218 (1974)\medskip 

\noindent
[Deser  1976a]
Nonlocal Conformal Anomalies, (with M.J. Duff and C.J. Isham),
Nucl.Phys. B111 (1976) 45 \medskip

\noindent
[Deser 1976b]
 Finite Energy Static Solutions to Chiral Models in 3 Space Dimensions, (with M. J. Duff and
C.J. Isham), Nucl. Phys. 114B, 29 (1976)\medskip

\noindent
[Deser 1976c]
Consistent Supergravity (with B. Zumino), Phys. Lett. 62B, 335 (1976).\medskip

\noindent
[Deser  1976d]
A Complete action for the Spinning String, 
(with B. Zumino), 
Phys.Lett. B65 (1976) 369 \medskip

\noindent
[Deser 1976e]
Duality Transformations of Abelian and Non-Abelian Gauge Fields (with C. Teitelboim), Phys. Rev. D13, 1592 (1976).\medskip

\noindent
[Deser  1977a]
 Renormalizability Properties of Supergravity, (with J.H. Kay and K.S. Stelle), Phys.Rev.Lett. 38 (1977) 527  \medskip

 \noindent
[Deser  1977b]
Supergravity Has Positive Energy (with C. Teitelboim), Phys. Rev. Lett. 39, 249 (1977). \medskip

\noindent
[Deser 1977c]
Broken Supersymmetry and Supergravity (with B. Zumino), Phys. Rev. Lett. 38, 1433 (1977).\medskip

\noindent
[Deser 1979]
Consistency Problems of Hypergravity (with C. Aragone), Phys. Lett. 86B, 161 (1979).\medskip

\noindent
[Deser  1980]
Gravitationally Induced CP Effects (with M.J. Duff and C.J. Isham), Phys. Letters 93B, 419 (1980).\medskip

\noindent
[Deser  1982]
Topologically Massive Gauge Theories, (with R. Jackiw and S. Templeton)
Annals Phys. 140 (1982) 372, \medskip

\noindent
[Deser  1984]
Three-Dimensional Einstein Gravity: Dynamics of Flat Space,
(with R. Jackiw and G. 't Hooft),  Annals Phys.152 (1984) 220. \medskip

 \noindent
 [Deser 1985b]
Inconsistency of Topologically Massive Hypergravity (with C. Aragone), Class. Quant. Grav. 2,
L29 (1985).\medskip

\noindent
[Deser  1993]
Geometric Classification of Conformal Anomalies in Arbitrary Dimensions (with A. Schwimmer), hep-th/9302042, Phys. Lett. B309, 279 (1993).\medskip

\noindent
[Deser 2021]
Forks In The Road: A Life In Physics
by Stanley Deser World Scientific (2021)
\medskip

 \section{\bf ACKNOWLEDGEMENTS}
%\newpage

MJD is very grateful to Chris Hull and Leron Borsten for discussions that led to improvements to the manuscript.
He would also like to thank Kelly Stelle's widow Marise Westbroek for permission to include Kelly's contributions.
%\newpage
\section{\bf AUTHOR PROFILES}

\begin{itemize}
\item{\bf Michael Duff}\\

Michael Duff (FRS 2009) is Emeritus Professor of Theoretical Physics and Distinguished Research Fellow at Imperial College London.  He was formerly Abdus Salam Professor of Theoretical Physics and Principal of the Faculty of Physical Sciences.
He is recipient of the 2004 Meeting Gold Medal, El Colegio Nacional, Mexico, the 2017 Paul Dirac Gold Medal and Prize, Institute of Physics, UK and the 2018 Trotter Prize,  USA.  He is a visitor to The Institute of Quantum Science and Engineering at Texas A\&M University and was inducted as 2019 Faculty Fellow of the Hagler Institute for Advanced Study there. A special issue of Proceedings of the Royal Society, https://royalsocietypublishing.org/topic/special-collections/duff-quantum-gravity,
Quantum Gravity, Branes and M-theory was dedicated him on the occasion of his 70th birthday.

\item{\bf Kellogg Stelle}\\ 

Kelly Stelle  was  Professor of Theoretical Physics at Imperial College London and formerly Head of the Theoretical Physics Group. He is the recipient of the 2006 Humboldt Research Award, Alexander von Humboldt Foundation, and the 2020 John William Strutt, Lord Rayleigh Medal and Prize, Institute of Physics. A tribute to his memory maybe found at https://www.online-tribute.com/KellyStelle.
\end{itemize}
  
  \begin{figure*}[h]
    \centering
    \begin{floatrow}\hspace{1cm}
      \ffigbox[6.5cm]{\caption*{Michael Duff FRS (Portrait by Beyond the Pixels.)}}{%
      \includegraphics[scale=0.165]{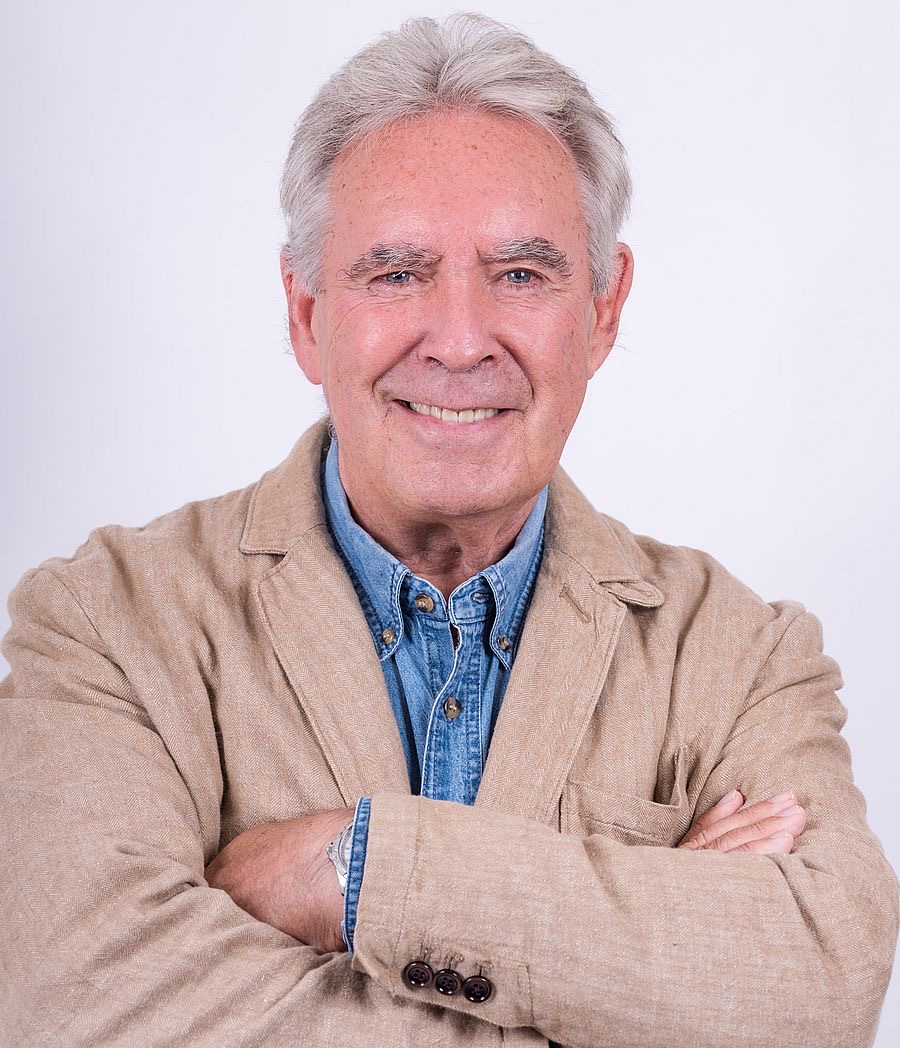} 
      }
      \hspace{1cm}
      \ffigbox[6.5cm]{\caption*{Kellogg Stelle (courtesy of Imperial College)}}{%
      \includegraphics[scale=0.60]{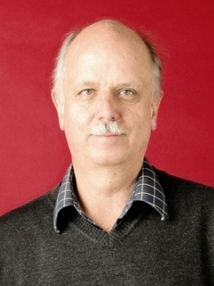}
      }
    \end{floatrow}
  \end{figure*}

\end{document}